\documentclass[12pt]{article}
\usepackage{graphicx}

\input{tcilatex}

\begin{document}

\title{Cosmic Crystallography of the Second Order}
\author{H. V. Fagundes and E. Gausmann \\
%EndAName
Instituto de F\'{\i}sica Te\'{o}rica, Universidade Estadual Paulista\\
Rua Pamplona, 145, S\~{a}o Paulo, SP, \\
CEP 01405-900, Brazil\\
Phone (5511) 31779090, Fax (5511) 31779080\\
E-mails: helio@ift.unesp.br, gausmann@ift.unesp.br}
\maketitle

\begin{abstract}
The cosmic crystallography method of Lehoucq et al. \cite{lelalu} produces
sharp peaks in the distribution of distances between the images of cosmic
sources. But the method cannot be applied to universes with compact spatial
sections of negative curvature. We apply to the these a second order
crystallographic effect, as evidenced by statistical parameters.


\end{abstract}

\section{Introduction}

The method of cosmic crystallography was developed by Lehoucq,
Lachi\`{e}ze-Rey, and Luminet \cite{lelalu}, and consists of plotting the
distances between cosmic images of clusters of galaxies. In Euclidean
spaces, we take the square of the distance between all pairs of images on a
catalogue versus the frequency of occurence of each of these distances. In
universes with Euclidean multiply connected spatial sections, we have sharp
peaks in a plot of distance distributions.

It is usual to consider the Friedmann-Lema\^{\i}tre-Robertson-Walker (FLRW)
cosmological models of constant curvature $(S^{3},E^{3},H^{3})$ with simply
connected spatial sections. However, models with these spacetime metrics
also admit, compact, orientable, multiply connected spatial sections, which
are represented by quotient manifolds $M=\tilde{M}/\Gamma $, where $\tilde{M}
$ is $S^{3}$, $E^{3}$ or $H^{3},$ and $\Gamma $ is a discrete group of
isometries (or rigid motions) acting freely and properly discontinously on $%
\tilde{M}$. The manifold $M$ is described by a fundamental polyhedron (FP)
in $\tilde{M}$, with faces pairwise identified through the action of the
elements of $\Gamma $. So $\tilde{M}$ is the universal covering space of $M$
and is the union of all cells $g($FP$)$, $g\in $ $\Gamma $.

The repeated images of a cosmic source is the basis of the cosmic
cristallography method. The images in a multiply connected universe are
connected by the elements $g$ of $\Gamma $. The distances between images
carry information about these isometries. These distances are of two types 
\cite{ULL990}: type I pairs are of the form $\{g(x),g(y)\},$ where \ 
\begin{equation}
distance[g(x),g(y)]=distance[x,y],
\end{equation}
for all points $x,y\in M$ and all elements $g\in $ $\Gamma $; type II pairs
of the form $\{x,g(x)\}$ if 
\begin{equation}
distance[x,g(x)]=distance[y,g(y)],  \label{clifford}
\end{equation}
for at least some points $x,y\in M$ and some elements $g$ of $\Gamma $.

\begin{figure}[tbh]
\centering\includegraphics[width=8cm]{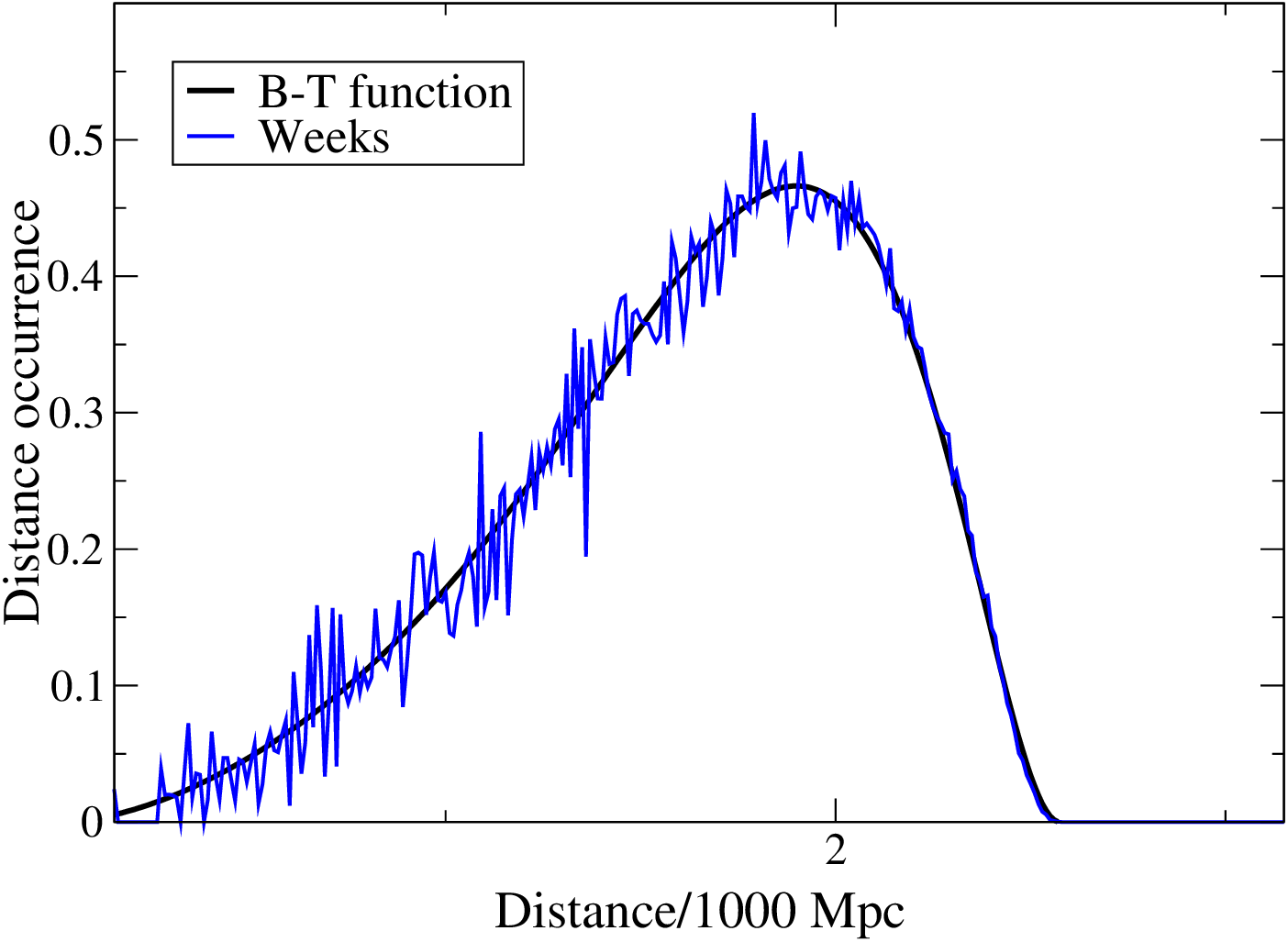}\newline
\caption{{\protect\small Distribution of pair distances of cosmic images for
the Bernui-Teixeira function, and for a universe with Weeks manifold as
spatial sections and }$\Omega _{0}=0.3.$}
\label{wt}
\end{figure}

The cosmic cristallography method puts in evidence type II pairs. These
distances are due to Clifford translations, which are elements $g\in \Gamma
, $ such that Eq. (\ref{clifford}) holds for \textit{any} two points $x,y\in
M. $ Type II pairs give sharp peaks in distance distributions in Euclidean 
\cite{lelalu,HG97} and spherical spaces \cite{GLLUW}, but they do not appear
in hyperbolic space. This is illustrated in Fig. \ref{wt} for an FLRW model
with total energy density $\Omega _{0}=0.3$ and having as spatial sections
the Weeks manifold - coded $m003(-3,1)$ in \cite{snapPea} and in Table I
below - which is the closed, orientable hyperbolic manifold with the
smallest volume (normalized to minus one curvature) known. The
Bernui-Teixeira (B-T) function \cite{bt99} is an analytical expression for a
uniform density distribution in an open hyperbolic model. \bigskip \bigskip\ 

\medskip

In hyperbolic spaces, the identity (or trivial motion) is the only Clifford
translation. In this case, the cosmic cristallography method by itself
cannot help us to detect the global topology of the universe. Several works
have tried to identify multiply connected, or closed, hyperbolic universes
by applying variants of the cosmic cristallographical method \cite{HG98,
HG99,ULL99,GRT2000}, most of which now rely on type I, in the absence of
type II, isometries. It is these variants that we call \textit{cosmic
crystallography of the second degree}.

One of these \cite{HG98}, proposed by us, consisted of subtracting, from the
distribution of distances between images in closed hyperbolic universes, the
similar distribution for the open model. It did not pretend to be useful for
the determination of a specific topology, but it might reinforce other
studies that look for nontrivial topologies.

Uzan, Lehoucq, and Luminet \cite{ULL99} invented the \textit{{collect
correlated pairs} }method, that collect type I pairs and plot them so as to
produce one peak in function of the density parameters, $\Omega _{m}$ for
matter and $\Omega _{\Lambda }$ for dark energy.

Gomero et al. \cite{GRT2000} obtained \textit{topological signatures, }by
taking averages of distance distributions for a large number of simulated
catalogues and subtracting from them averages of simulations for trivial
topology.

Here we introduce still another second order crystallographic method, in the
absence of Clifford translations and sharp peaks. We look for signals of
nontrivial topology in statistical parameters of their distance
distributions.

As commented above on Ref. \cite{HG98}, these methods are not as powerful as
the original Clifford crystallography, but will certainly be useful as added
tools to help looking for the global shape of the universe.

\section{Simulations}

Let the metric of the Friedmann's open model be written as

\begin{equation}
ds^{2}=a^{2}(\eta )(d\eta ^{2}-d\lambda ^{2})\ ,
\end{equation}
where $a(\eta )$ is the expansion factor or curvature radius, and 
\begin{equation}
d\lambda ^{2}=d\chi ^{2}+\sinh ^{2}\chi (d\theta ^{2}+\sin ^{2}\theta
\,d\phi ^{2})
\end{equation}
is the standard metric of hyperbolic space $H^{3}$. We assume a null
cosmological quantity, hence the expressions for $a(\eta )$ and other
quantities are as in Friedmann's open model - see, for example, Landau and
Lifshitz \cite{landau}.

To simulate our catalogues we assume for the cosmological density parameter
the values $\Omega _{0}=0.3$ and $\Omega _{0}=0.4,$ with Hubble's constant $%
H_{0}=65\,$km\thinspace s$^{-1}$Mpc$^{-1}$. The present value of the
curvature radius is $a(\eta _{0})=5513$ Mpc for $\Omega _{0}=0.3$ and $%
5954\, $Mpc for $\Omega _{0}=0.4$.

To generate pseudorandom source distributions in the FP, we first change the
coordinates to get a uniform density in coordinate space: 
\[
dV=\sinh ^{2}\chi \sin \theta \,\,d\chi \,d\theta \,d\phi =du\,dv\,d\phi \,, 
\]
with $u(\chi )=(\sinh \chi \cosh \chi -\chi )/2$ and $v(\theta )=\cos
^{-1}\theta $. Our sources are then generated with equal probabilities in $%
(u,v,\phi )$ space, and their large scale distributions are spatially
homogeneous. \ 

\medskip 
\begin{tabular}{|cccc|cc|}
\multicolumn{6}{l}{\small Table I. The eight manifolds used as spatial
sections. Their names and} \\ 
\multicolumn{6}{l}{\small data are taken from the SnapPea software. The last
two columns give} \\ 
\multicolumn{6}{l}{{\small a} {\small number of cells that cover the
observable universe.}} \\ \hline\hline
\multicolumn{4}{|c|}{Manifold} & \multicolumn{2}{|c|}{Number of cells} \\ 
\hline
Name & Volume & $\chi _{in}$ & $\chi _{out}$ & $\Omega _{0}=0.3$ & $\Omega
_{0}=0.4$ \\ \hline
\multicolumn{1}{|l}{m003(-3,1)} & 0.94 & 0.52 & 0.75 & 747 & 379 \\ 
\multicolumn{1}{|l}{m003(-2,3)} & 0.98 & 0.54 & 0.75 & 729 & 357 \\ 
\multicolumn{1}{|l}{m017(-3,2)} & 1.89 & 0.64 & 0.85 & 463 & 247 \\ 
\multicolumn{1}{|l}{m221(+3,1)} & 2.83 & 0.74 & 0.94 & 403 & 199 \\ 
\multicolumn{1}{|l}{m342(-4,1)} & 3.75 & 0.77 & 1.16 & 451 & 237 \\ 
\multicolumn{1}{|l}{s890(+3,2)} & 4.69 & 0.87 & 1.38 & 653 & 273 \\ 
\multicolumn{1}{|l}{v2051(+3,2)} & 4.69 & 0.87 & 1.38 & 653 & 273 \\ 
\multicolumn{1}{|l}{v2293(+3,2)} & 4.69 & 0.87 & 1.38 & 653 & 273 \\ 
\hline\hline
\end{tabular}
\bigskip \bigskip

We did the simulations for eight spatially compact, hyperbolic models. Their
space sections are the manifolds listed in Table I, which gives their names,
volumes, and the circumscribing and inscribing radii ($\chi _{out}$ and $%
\chi _{in}$) of their FP's. These and other data on the manifolds were
obtained from the SnapPea program \cite{snapPea}. In the last two columns,
we list the number of cells (replicas of FP) needed to assure a complete
cover \ of $H^{3}$ up to a radius $\chi =2.335$ for $\Omega _{0}=0.3$ and $%
\chi =1.996$ for $\Omega _{0}=0.4$. This corresponds to the redshift ($%
Z=1300 $) of the last scattering surface. Aproximately $850$ images were
created in each simulated catalogue. Manifolds with different volumes will
have different numbers of sources in their FP's.\medskip \bigskip 
\begin{figure}[tbp]
\centering\includegraphics[width=8cm]{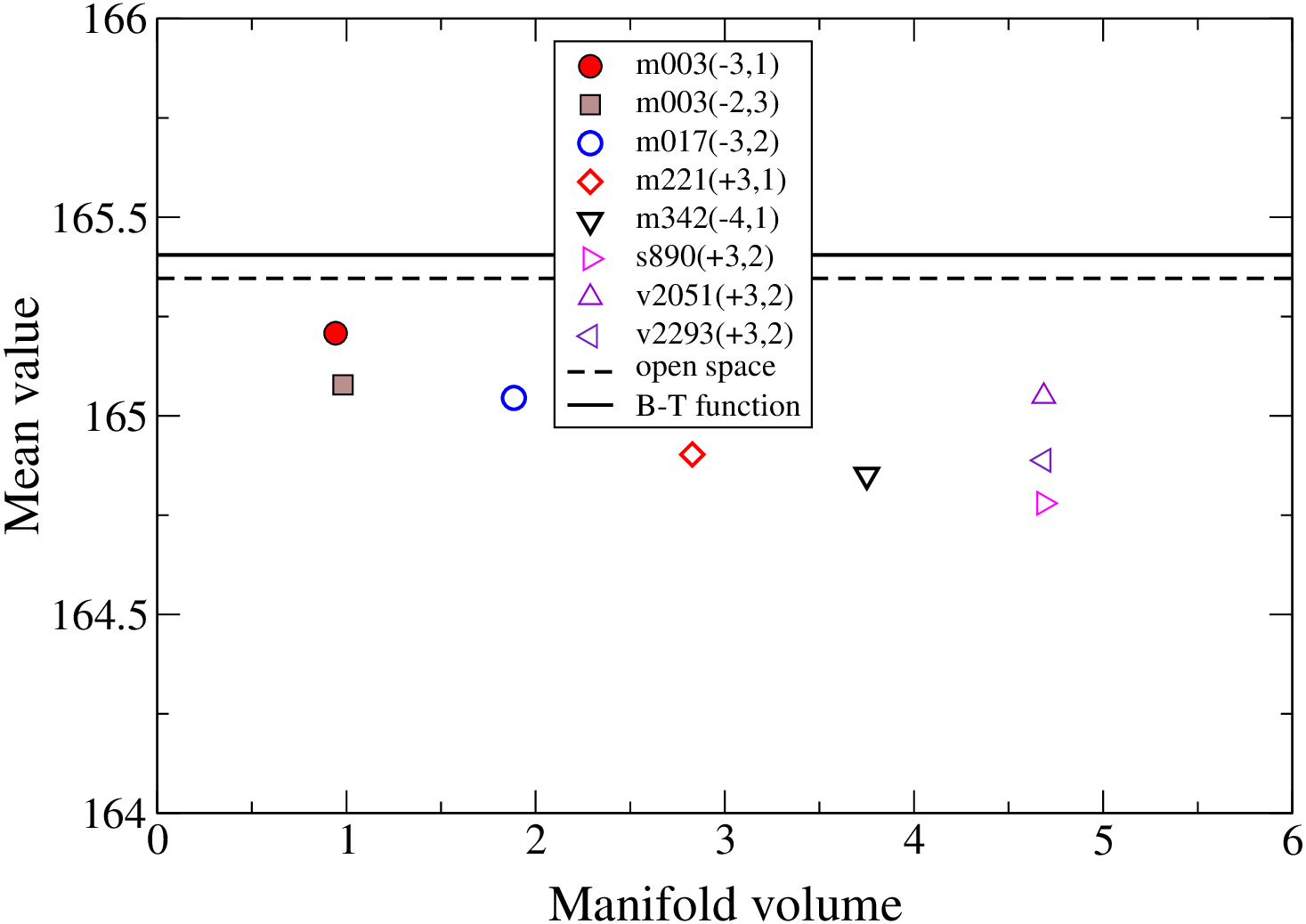}\newline
\caption{{\protect\small The mean values for the eight studied manifolds,
with }$\Omega _{0}=0.3.${\protect\small \ In this and remaining figures, a
dashed line marks the value of the plotted parameter for the open model, and
a solid line marks the Bernui-Teixeira value. }}
\label{5vm}
\end{figure}
\ 

\begin{figure}[tbp]
\centering\includegraphics[width=8cm]{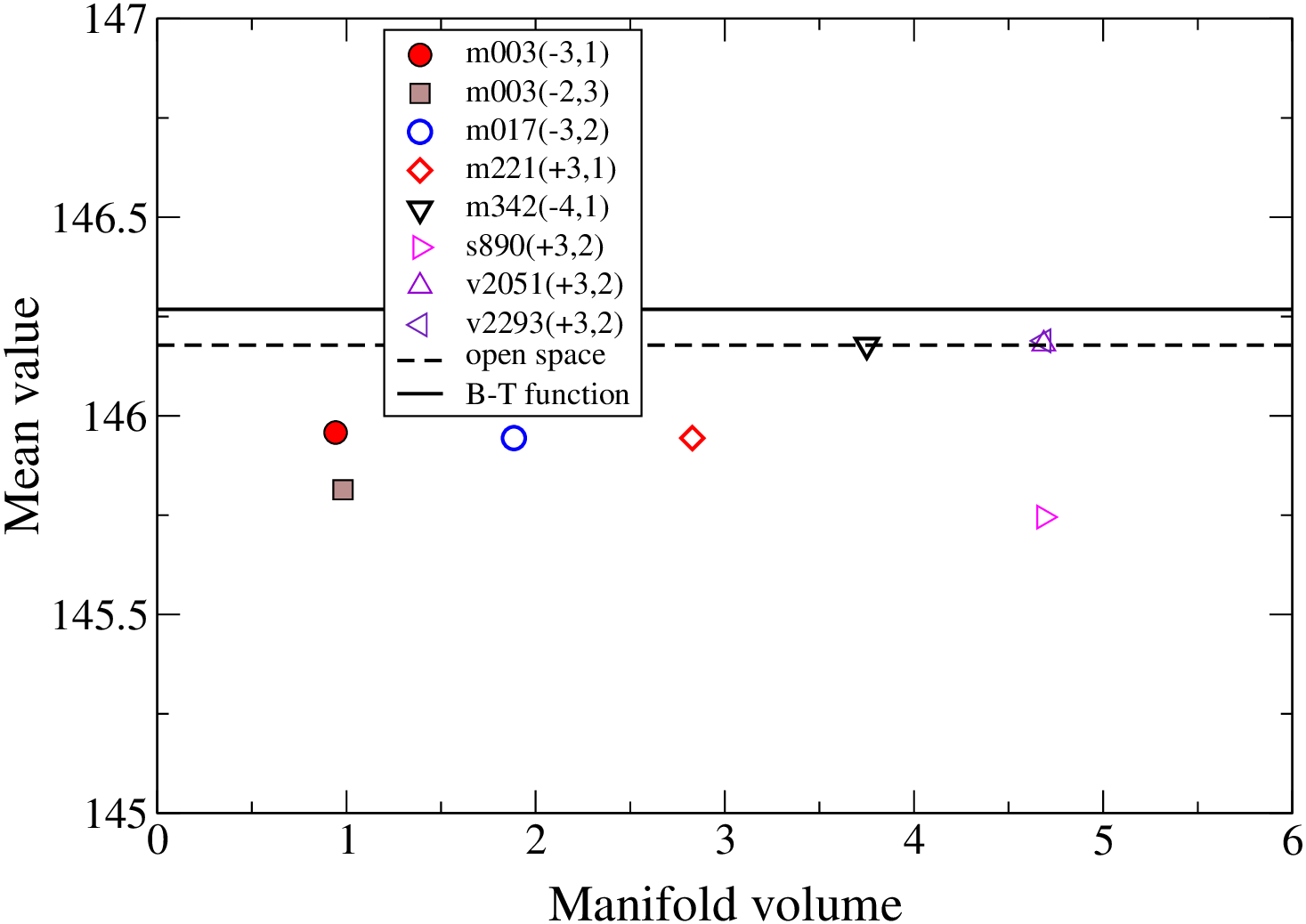}\newline
\caption{{\protect\small The mean values for our simulations with }$\Omega
_{0}=0.4${\protect\small .}}
\label{2vm}
\end{figure}

%Fig. 4
\begin{figure}[tbp]
\textit{\centering\includegraphics[width=8cm]{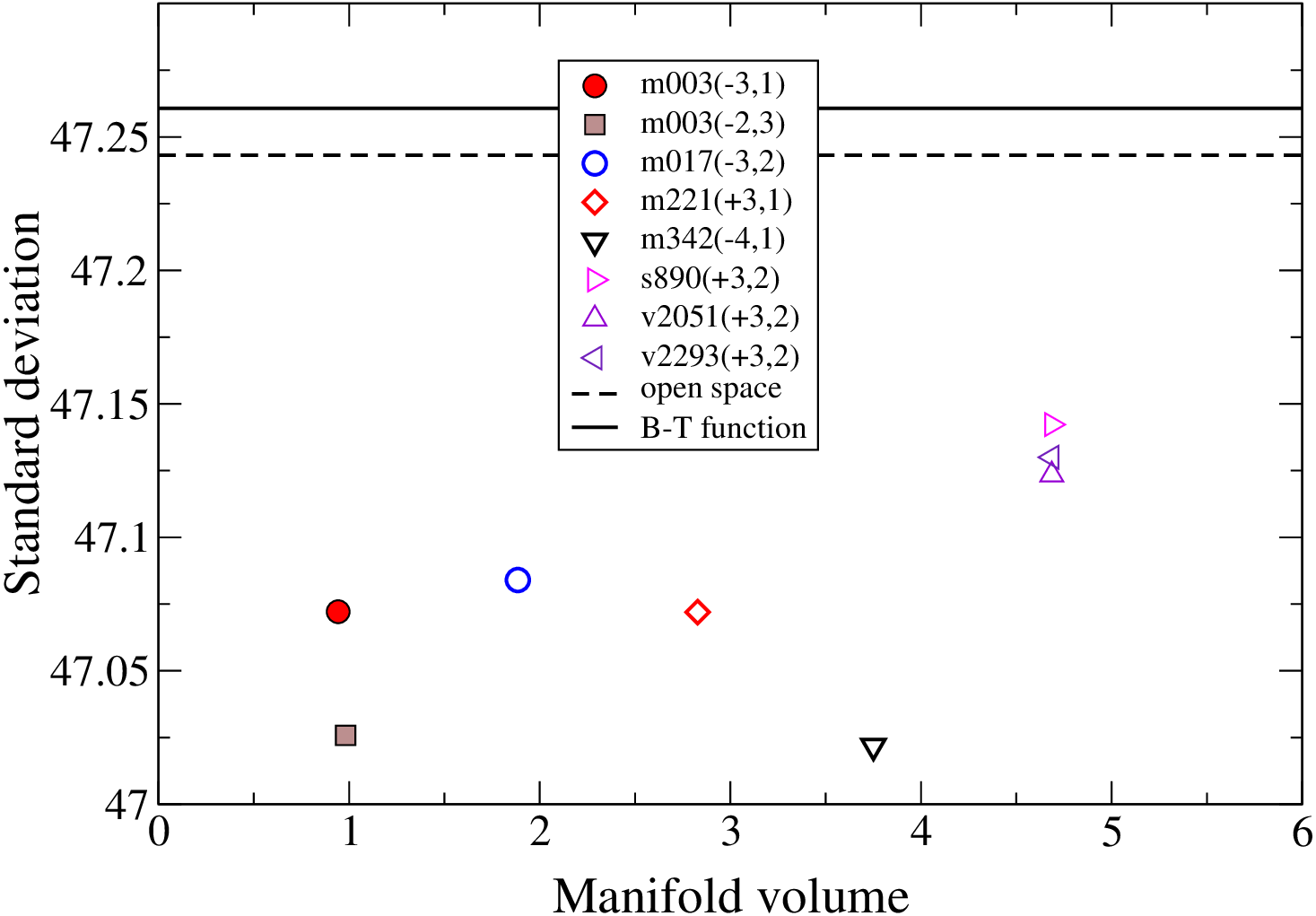}\newline
}
\caption{{\protect\small The standard deviations }$\protect\sigma $%
{\protect\small \ for the simulations with }$\Omega _{0}=0.3.$}
\label{5dp}
\end{figure}

\bigskip

%Fig. 5
\begin{figure}[tbp]
\textit{\centering\includegraphics[width=8cm]{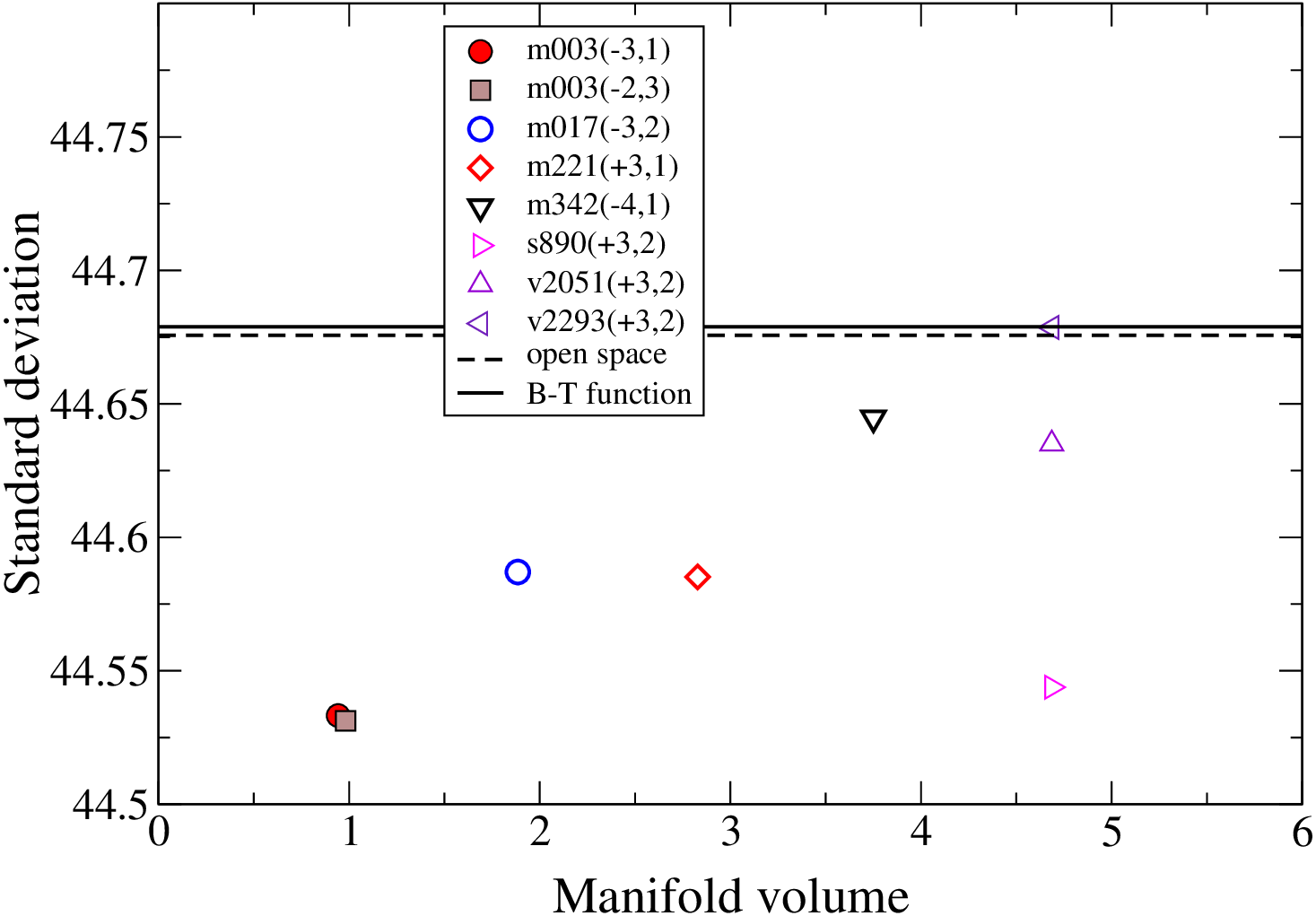}\newline
}
\caption{{\protect\small The standard deviations for }$\Omega _{0}=0.4.$}
\label{2dp}
\end{figure}

%Fig. 6
\begin{figure}[tbp]
\textit{\centering\includegraphics[width=8cm]{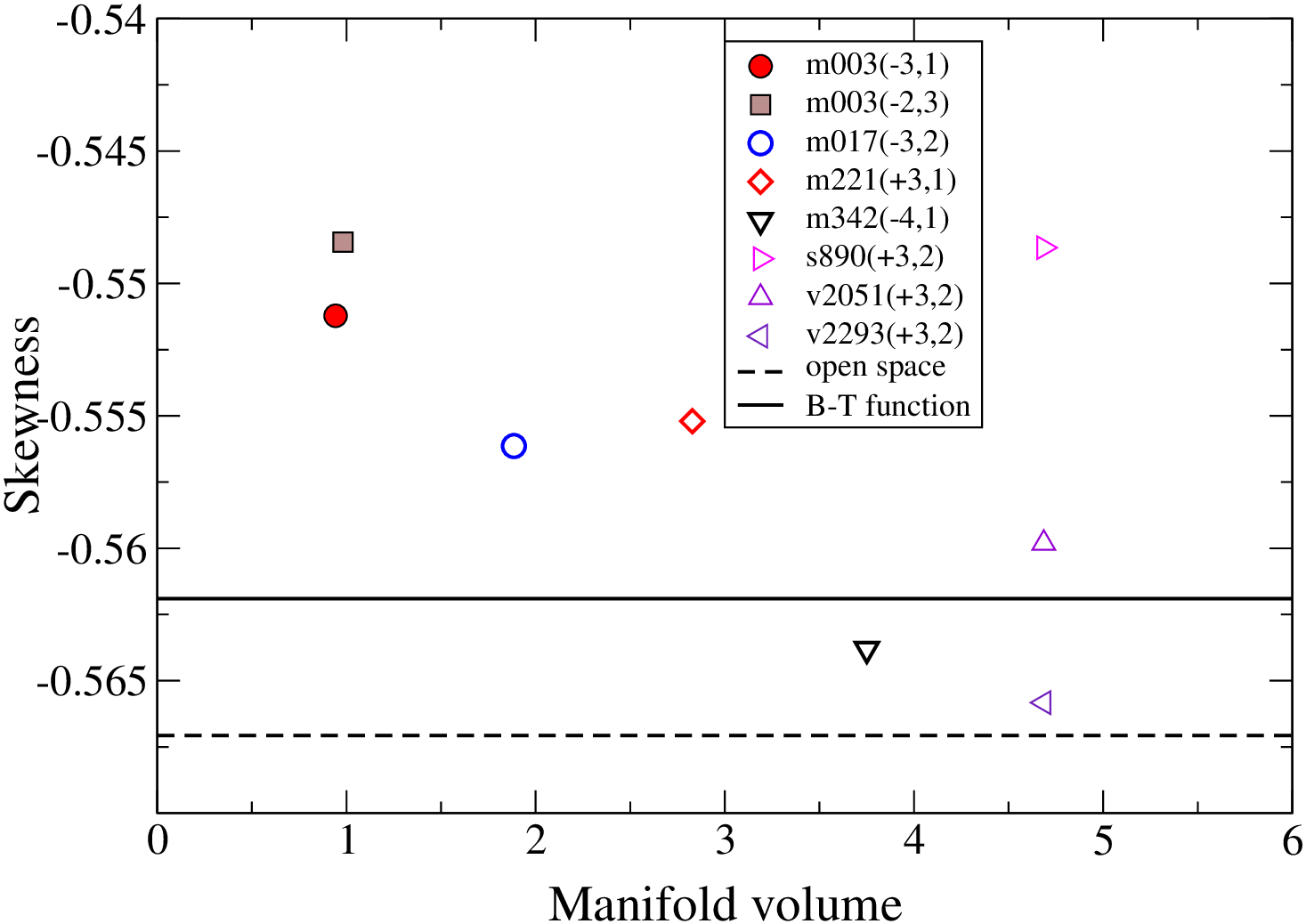}\newline
}
\caption{{\protect\small The skewnesses for }$\Omega _{0}=0.3$%
{\protect\small .}}
\label{5sn}
\end{figure}

%Fig. 7
\begin{figure}[tbp]
\textit{\centering\includegraphics[width=8cm]{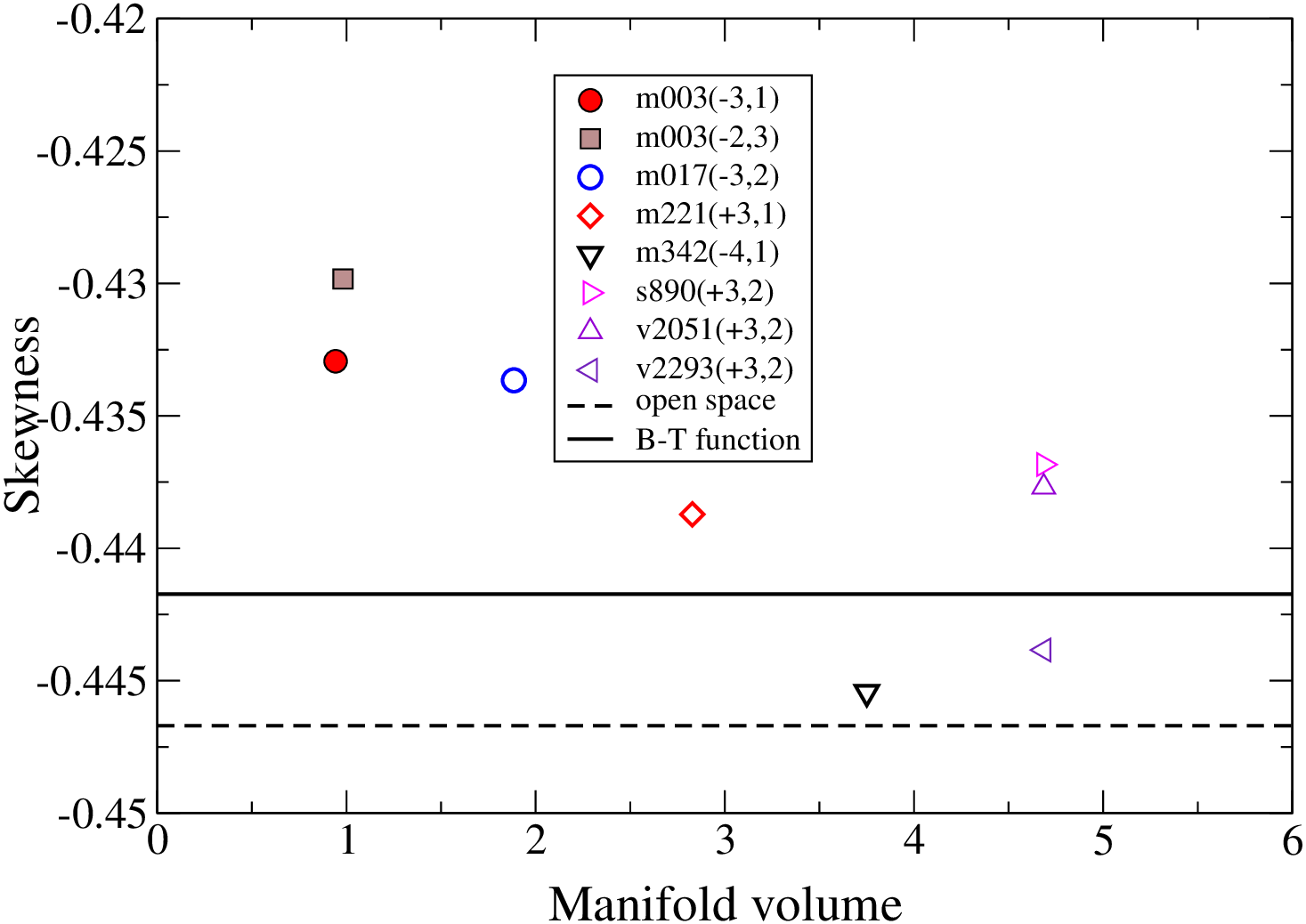}\newline
}
\caption{{\protect\small The skewnesses for }$\Omega _{0}=0.4$%
{\protect\small .}}
\label{2sn}
\end{figure}

\bigskip \bigskip

%Fig. 8
\begin{figure}[tbp]
\textit{\centering\includegraphics[width=8cm]{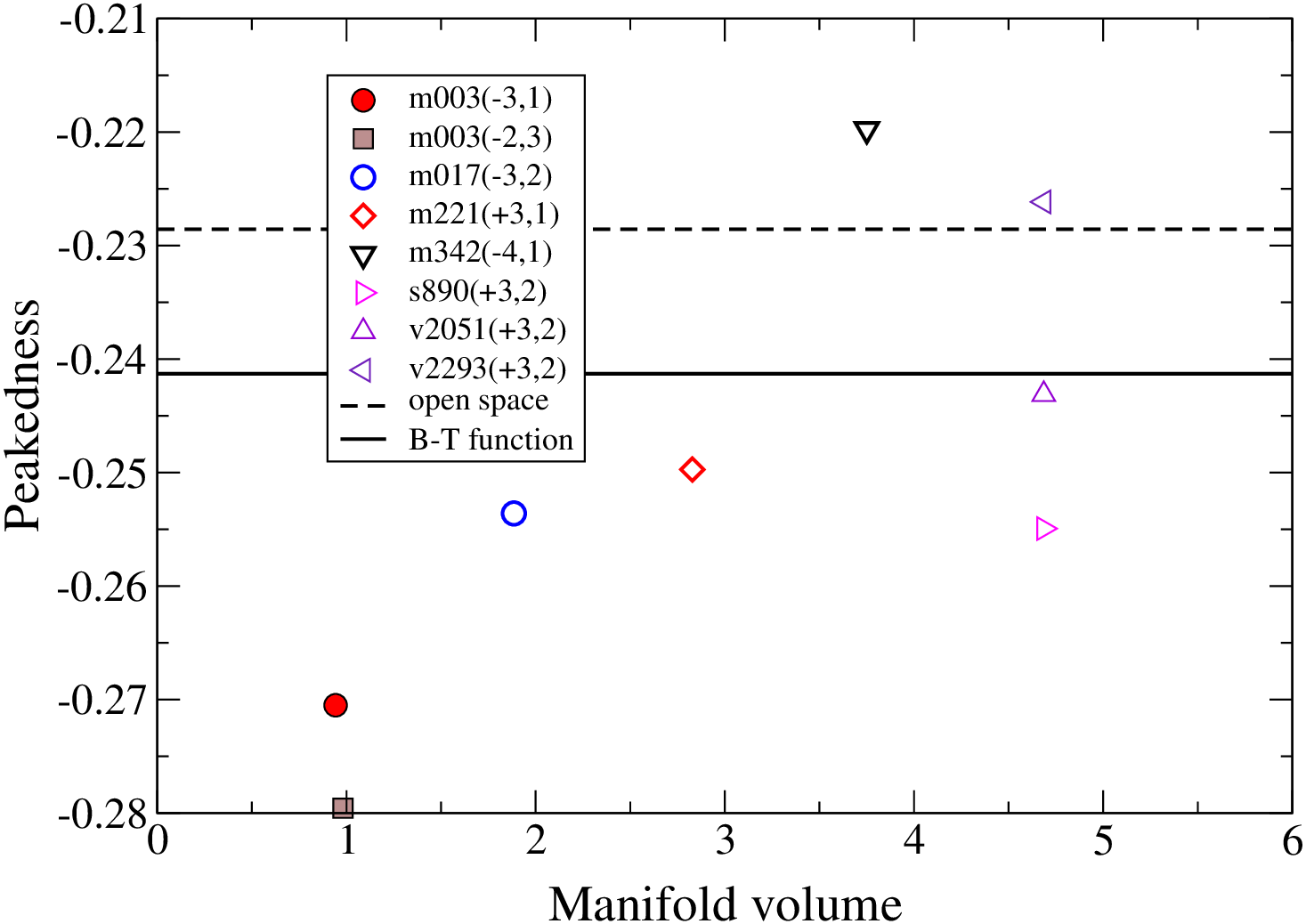}\newline
}
\caption{{\protect\small The peakednesses for }$\Omega _{0}=0.3.$}
\label{5pd}
\end{figure}

%Fig. 9
\begin{figure}[tbp]
\textit{\centering\includegraphics[width=8cm]{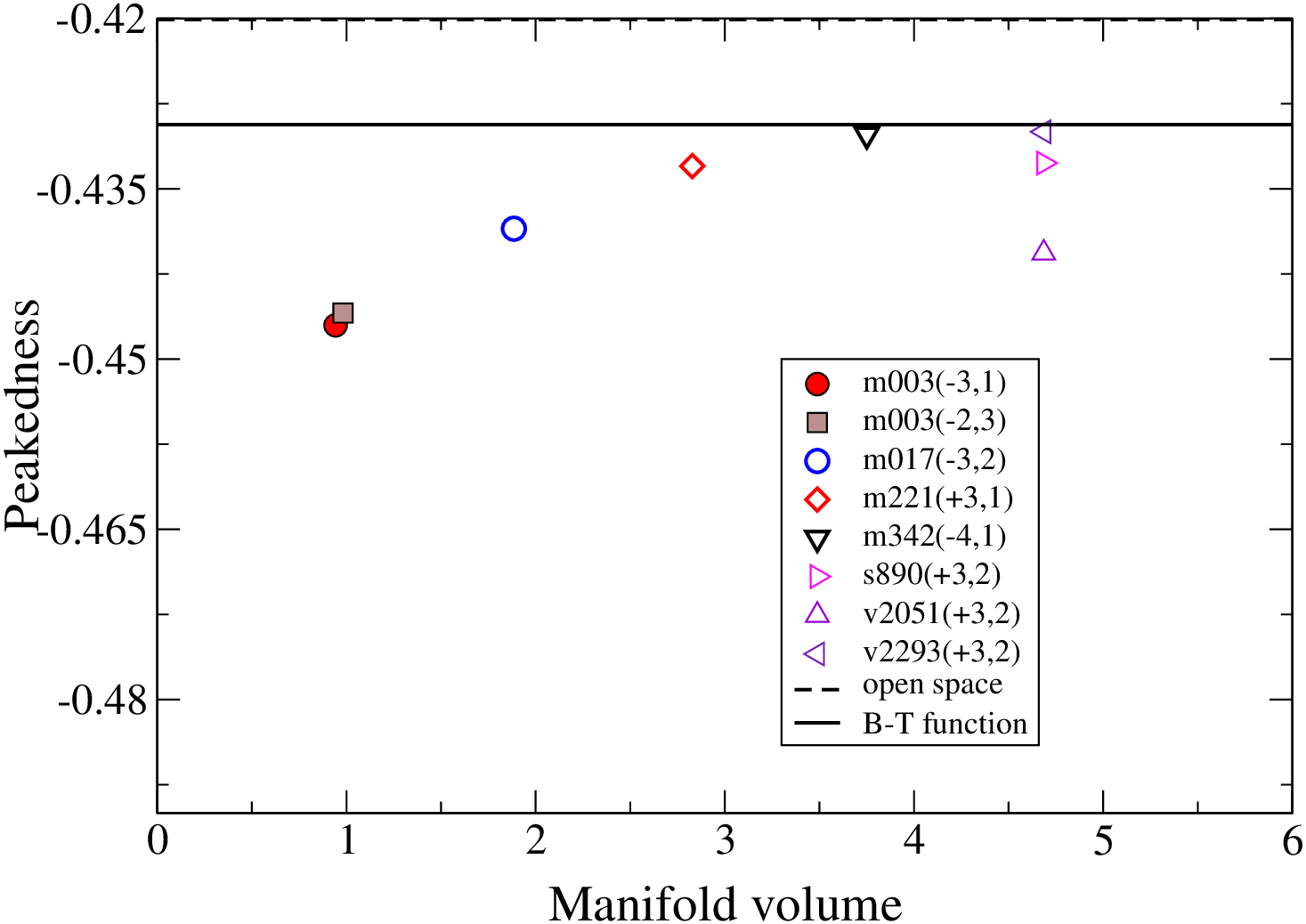}\newline
}
\caption{{\protect\small The peakednesses for }$\Omega _{0}=0.4.$}
\label{2pd}
\end{figure}
\bigskip

Ten different pseudorandom distributions of sources for each manifold were
simulated. From the result of each simulation we calculated the distance
distribution and then the latter's mean value $mv=E[X],$ standard deviation $%
\sigma =\sqrt{\mu _{2}},$ skewness\footnote{%
Skewness characterizes the degree of asymmetry of a distribution around its
mean value. A negative value means a leftward asymmetry (as in Fig. 1), a
positive value a rightward one.} $sn=\mu _{3}/\sigma ^{3},$ and peakedness%
\footnote{%
Or \textit{kurtosis}, which measures the relative peakedness or flatness of
a given distribution, relative to the normal distribution.} $pn=(\mu
_{4}/\sigma ^{4})-3,$ where $X$ is the distance variable, $E$ is the
expected value operator, and $\mu _{\nu }=E[(X-mv)^{\nu }]$ is the $\nu 
$-moment about the mean value - cf. \cite{barnes}, for example. 

For each of our chosen eight compact manifolds, we took the averages of
these statistical parameters for ten simulations (obtained by varying the
computer pseudorandom seed), to get their tendency in comparison with those
for the simply connected case. \medskip

We proceeded in two ways to obtain a distribution in the open model, for
comparison with the compact cases. In one of them, a simulated catalogue
(real ones are not yet deep enough for use in cosmic crystallography) was
obtained, with the same cosmological parameters and a pseudorandom
distribution of sources inside the observable universe (redshift $\leq 1300$%
). In the other, the analytical Bernui-Teixeira function for a uniform
distribution in simply connected universes \cite{bt99} was used.

\section{Results and conclusions}

We compare the results for eight compact manifolds with those for the
simulated simply connected case, and with the B-T function. Figures 2-9 show
these results for $mv$, $\sigma $, $sn$, and $pn$, plotted vs. the manifold
volumes, with the density parameters $\Omega _{0}=0.3$ and $\Omega _{0}=0.4$%
. It is of course premature to think of a functional dependence on volume
(except for the obvious fact that, if the volume is so large as to enclose
the observable universe, then the distribution of sources is
indistinguishable from that for an open universe - see Fagundes \cite{HVF83}%
, for example). But we note a tendency of $mv$, $\sigma $, and $pn$ in
compact universes to have values below those of the corresponding B-T
function, and for $sn$ to lie above that function. This could give us a
signal of nontrivial topology in a future, realistic situation.

We expect that the presence of type I pairs of distances is more evident for
manifolds with smaller normalized volumes, or for models with smaller $%
\Omega _{0}$. In fact our results are better for $mv$ and $\sigma $ with $%
\Omega _{0}=0.3$, where the observable universe is bigger than for $\Omega
_{0}=0.4$. This implies more copies of the FP, and hence more topological
effects. The set of these statistical parameters may eventually provide a
complementary indication as for the multiply connectedness of a possibly
negatively curved cosmos.\bigskip\ \ 

\bigskip

E. G. is grateful to Funda\c{c}\~{a}o de Amparo \`{a} Pesquisa do Estado de
S\~{a}o Paulo (FAPESP Process 01/10328-6) for a post-doctoral scholarship.
H. V. F. thanks Conselho Nacional de Desenvolvimento Cient\'{\i}fico e
Tecnol\'{o}gico (CNPq Process 300415/84-2) for partial financial support.

%\begin{table}
%  \centering
%  \begin{tabular}{|c|c|c|c|c|c|}
%   x & VM & $\sigma$ & sn & pn & cv \\
%   med weeks& 152.8651 & 46.791142 & 0.235478 & 2.430151 & 30.609568\\
%   med thurs& 152.866 & 46.774568 & 0.234415 & 2.430817 & 30.598538 \\
%  med s890& 152.5472 & 46.8565114 & 0.233048 & 2.451136 & 30.716201 \\
%  med v2051& 152.7967 & 46.843758 & 0.224045 & 2.45057 & 30.657933 \\
%  med v2293& 152.6518 & 46.847847 & 0.21754 & 2.461198 & 30.689774 \\
%  med ran& 153.0709 & 46.960801 & 0.2167 & 2.457232 & 30.680885 \\
%  Teix & 153.126 & 46.97794 & .22150 & 2.45057 & 30.67933 \\
% \end{tabular}
%  \caption{omega $0.3$}\label{4}
%\end{table}

%\begin{table}
% \centering
% \begin{tabular}{|c|c|c|c|c|c|}
%   x & VM & $\sigma$ & sn & pn & cv \\
%   med weeks& 145.8516 & 44.542196 & -0.4324 & 2.558063 & 30.539558 \\
%   med thurs& 145.7585 & 44.486214 & -0.433377 & 2.556513 & 30.520612 \\
%   med s890& 145.7458 & 44.543907 & -0.436837 & 2.567264 & 30.562815 \\
%   med v2051& 146.1819 & 44.635257 & -0.437706 & 2.559372 & 30.534229 \\
%   med v2293& 146.189 & 44.678763 & -0.44385 & 2.570044 & 30.562381 \\
%   med ran& 146.178 & 44.675856 & -0.446714 & 2.579922 & 30.564281 \\
%   Teix & 146.269 & 44.67937 & -0.44175 & 2.57069 & 30.54597 \\
%  \end{tabular}
% \caption{omega $0.4$}\label{14}
%\end{table}

%\newpage

\end{document}